\pdfoutput=1
\documentclass[sigconf,nonacm,screen,balance=false]{acmart}
\usepackage{popets}
\usepackage{accents}
\usepackage[noend,ruled,linesnumbered]{algorithm2e}
\usepackage{microtype}
\usepackage{multirow}
\usepackage{pgfplots}
\usepackage{standalone}
\usepackage{tikz}
\usepackage{amsthm}
\usepackage{hyperref}
\usepackage{todonotes}
\usepackage{pifont}
\usepackage{subcaption}
\PassOptionsToPackage{numbers}{natbib}

\newcommand{\xmark}{\ding{55}}%

\newtheorem{definition}{Definition}

\pgfplotsset{compat=1.18}
\usetikzlibrary{
  arrows, calc, chains,
  decorations.pathreplacing,
  matrix, positioning, scopes,
  shapes.arrows, shapes.callouts,
  shapes.multipart,
}

\RestyleAlgo{ruled}

\setcopyright{popets}
\copyrightyear{2024}

\acmYear{YYYY}
\acmVolume{YYYY}
\acmNumber{X}
\acmDOI{XXXXXXX.XXXXXXX}
\acmISBN{}
\acmConference{Proceedings on Privacy Enhancing Technologies}
\begin{document}


\title[Homomorphic WiSARDs]{
  Homomorphic WiSARDs: Efficient Weightless Neural
  Network training over encrypted data
}


\author{Leonardo Neumann}
\orcid{0000-0002-4409-6118}
\affiliation{%
  \institution{University of Campinas}
  \city{Campinas}
  \state{São Paulo}
  \country{Brazil}
} 
\email{leonardo@neumann.dev.br}

\author{Antonio Guimarães}
\orcid{0000-0001-5110-6639}
\affiliation{%
  \institution{IMDEA Software Institute}
  \city{Madrid}
  \state{}
  \country{Spain}
}
\email{antonio.guimaraes@imdea.org}

\author{Diego F. Aranha}
\orcid{0000-0002-2457-0783}
\affiliation{%
  \institution{Aarhus University}
  \city{Aarhus}
  \state{}
  \country{Denmark}
}
\email{dfaranha@cs.au.dk}

\author{Edson Borin}
\orcid{0000-0003-1783-4231}
\affiliation{%
  \institution{University of Campinas}
  \city{Campinas}
  \state{São Paulo}
  \country{Brazil}
}
\email{borin@unicamp.br}


\renewcommand{\shortauthors}{Neumann et al.}


\begin{abstract}

  The widespread application of machine learning algorithms is a matter of
  increasing concern for the data privacy research community, and many have
  sought to develop privacy-preserving techniques for it. Among the existing
  approaches, the homomorphic evaluation of machine learning algorithms stands
  out by performing operations directly over encrypted data, enabling strong
  and inherent guarantees of confidentiality. The HE evaluation of inference
  algorithms is already practical even for relatively deep Convolution Neural
  Networks (CNNs). However, training is still a major challenge, with current
  solutions often resorting to interactive protocols or lightweight 
  algorithms, which can be unfit for accurately solving more complex problems,
  such as image recognition.

  In this work, we introduce the homomorphic evaluation of Wilkie, Stonham, and
  Aleksander’s Recognition Device (WiSARD) and subsequent state-of-the-art
  Weightless Neural Networks (WNNs) both for training and inference on
  homomorphically encrypted data. Compared to CNNs, WNNs offer much better
  performance with a relatively small accuracy deterioration. We develop a
  complete framework for it, including several building blocks that can be of
  independent interest. Our framework achieves 91.71\% accuracy on the MNIST
  dataset after only 3.5 minutes of encrypted training (multi-threaded), going
  up to 93.76\% in 3.5 hours after training over 60 thousand images. For the
  HAM10000 dataset, we achieve 67.85\% accuracy in just 1.5 minutes, going up
  to 69.85\% after 1 hour. Compared to the state of the art on HE evaluation of
  CNN training, Glyph (Lou \emph{et~al.}, NeurIPS 2020), these results
  represent a speedup of up to 1200 times with an accuracy loss of at
  most 5.4\%. For HAM10000, we even achieved a 0.65\% accuracy improvement
  while being 60 times faster than Glyph. We also provide solutions for small-scale encrypted training. In a single thread on a consumer Desktop machine 
  using less than 200MB of memory, we train over 1000 MNIST images in 12
  minutes or over the entire Wisconsin Breast Cancer dataset in just 11
  seconds.

\end{abstract}



\maketitle


\section{Introduction}

The popularization of machine learning (ML) in modern data processing
applications brought with itself a great concern over the privacy implications
of its widespread use, which often requires large-scale data collection or
processing of sensitive information. As a result, privacy-preserving machine
learning became a topic of broad research interest, and many solutions have
been proposed on different fronts of this issue. Among them, homomorphic encryption (HE) has long been
considered one of the most powerful tools for enabling privacy for data during
processing, as it enables operations to be performed directly over encrypted
data. It comes, however, with a significant computational performance overhead,
which has often been an impairment for the HE evaluation of large ML models.

The most successful use cases so far are in the HE evaluation of inference algorithms based on Convolutional Neural Networks (CNNs)~\cite{hesamifard:2017,giladbachrach:2016,izabachene:2019}. Thanks to techniques such as quantized training~\cite{legiest:2023} and FHE-friendly activation functions~\cite{ishiyama:2020}, modern HE schemes are capable of efficiently evaluating inferences while achieving near state-of-the-art accuracy. Efficient and accurate encrypted training, on the other hand, remains mostly an open problem, as current solutions either stay far from state-of-the-art accuracy levels or take weeks of computation to be run~\cite{lou:2020}. Some scenarios allow for alternative approaches, such as client-assisted FHE~\cite{brand:2023} and other MPC-based protocols. These commonly offer good results in limited contexts, but they come with the intrinsic downsides of a multi-party protocol. 

Other solutions rely on techniques such as transfer learning and cleartext
feature extraction, in which a pre-trained model or feature extraction
algorithm is applied over the unencrypted input data before the encrypted
training. This approach has often been employed to accelerate or improve the
accuracy of encrypted training~\cite{lou:2020,lee:2023,panzade:2024}. However,
its application is limited as it requires the existence of a pre-trained model
that works for the specific problem. This model needs to be trained and
evaluated on cleartext, requiring additional security and usability
assumptions. It requires, for example, the client (data owner) to perform part
of the training (the feature extraction) on cleartext in a trusted environment
before encrypting the input data, which might be prohibitive in the case of
resource-constrained clients. 

For end-to-end fully encrypted training, the most practical approaches so far
go in the direction of adopting simpler ML algorithms or methods for switching
between different HE schemes. In 2019, NRPH19~\cite{nandakumar:2019} presented
the first stochastic-gradient-descent-based (SGD) approach for encrypted NN
training, achieving accuracies of 85.9\% up to 97.8\% on inferences on the
MNIST dataset~\cite{lecun:2010}. Their execution time, however, would be up to
more than a decade (as estimated in~\cite{lou:2020}). Using HE scheme switching,
Chimera~\cite{boura:2018,boura:2019} and Glyph~\cite{lou:2020}, improved this
result significantly, enabling accuracies of 94.1\% up to 97.1\% with execution 
times between 5.7 and 28.6 days\footnote{From Glyph results.} for the former
and between 1.5 and 8 days for the latter. Recently, 
MFK$^+$24~\cite{montero:2024} presented a
TFHE~\cite{chillotti:2020}-based evaluation of a multi-layer perceptron (MLP)
training for the Wisconsin breast cancer dataset~\cite{wolberg:1995}, achieving
98.25\% accuracy in 49 minutes.


\subsection{Overview and Contributions}

In this work, we follow the general idea of evaluating alternative NN models to
introduce the HE evaluation of Weightless Neural Networks (WNNs). In contrast
with typical NNs, neurons of WNNs are Random Access Memory units (RAMs), which
evaluate arbitrary functions over the data they receive. There are no weights
or biases, and the training consists of programming the RAM units to evaluate
different functions that will (hopefully) recognize categorizing patterns in
the input data. One of the most basic instantiations of a WNN and our starting
point in this work is Wilkie, Stonham, and Aleksander’s Recognition Device
(WiSARD)~\cite{aleksander:1984}. In their model, RAM units are programmed to
simply output whether or not a certain pattern of bits (the neuron's input) was
present in the training set for a certain label. Modern WNNs employ
significantly more complex functions than this, but the basic principle of
programming RAM units remains their defining aspect.

Our work starts from the observation that operating with RAM units fits
perfectly well within the HE evaluation model provided by schemes such as
FHEW~\cite{ducas:2015} and TFHE~\cite{chillotti:2020}, which are based on the
evaluation of lookup tables (LUTs).

\subsubsection{Evaluation model}
The main aspect determining the characteristics of a WNN is the type of RAM
it implements. The original WiSARD model adopted binary RAMs,
\emph{i.e.}, each address of the RAM stores a single bit, and, hence, the RAM
of each neuron can only indicate whether a pattern occurred or not. Subsequent
WNNs adopted a \textit{bleaching technique} that uses integers as RAM elements
followed by a threshold function. In this case, each address counts the number
of occurrences of a pattern and, at the end of the training, a threshold
function binarizes the elements based on some fixed value. Modern WNNs employ a
variety of techniques to implement RAMs, including, \emph{e.g.},
Bloom filters and SGD-adjusted scores.

In this work, our first step is to represent WNNs in a model that can be
efficiently homomorphically evaluated. For this, we generalize the bleaching
technique approach and formalize the concepts of the Integer WiSARD model
and model activation phases. The basic working principle remains the same, we
use integer RAMs to count every occurrence of a pattern and apply a non-linear
function over the RAMs once training is finished. The main differences lie in
the separation of the process into two procedures and in the generalization of
bleaching to allow the performing of arbitrary activation functions. Our main
motivation for this is to separate arithmetic (integer counting) from
non-arithmetic (model activation) computation, which not only facilitates the
HE evaluation but also enables us to introduce several optimizations of
independent techniques for each procedure. Nonetheless, our approach also
brings advantages for the evaluation of WNNs in general. The use of activation
functions other than just threshold binarization allows us to implement other
types of WNNs following the same principle. It also allows us to consider
alternative activation functions to improve the average accuracy. Additionally,
the Integer WiSARD model training, as a standalone procedure, is trivially
parallelizable and enables us to easily define protocols for distributed and
federated learning.


\subsubsection{Encrypted Model Inference}

We introduce the HE evaluation of WNN inference algorithms using the TFHE
scheme~\cite{chillotti:2020}.
Compared to other existing practical solutions for NN inference over encrypted
data, our approach has the advantage of providing model privacy, as our
evaluation method can work with an encrypted model with minimal additional
cost. It also significantly advances the state of the art on TFHE-based NN
inference~\cite{stoian:2023}. 

TFHE introduced procedures such as programmable
bootstrapping~\cite{chillotti:2021} and vertical packing~\cite{chillotti:2020},
which enable the evaluation of arbitrary functions by representing them as 
lookup tables. Thanks to this, the scheme is often highlighted as one of the 
most promising schemes for NN applications. Realizing this potential, however,
has shown to be a challenge, as most of its current applications struggle to
provide low inference latency due to the combination of arithmetic and
non-arithmetic procedures of typical NN. Our evaluation model for WNNs avoids
this problem, and we improve the performance of state-of-the-art TFHE-based
inference~\cite{stoian:2023} more than 300 times.


\subsubsection{Encrypted Model Training}

Our main focus and contribution is the introduction of a framework to allow the
efficient homomorphic evaluation of WNN training. We perform training encrypted
end to end, without using any pre-processing techniques that would require
knowledge of the full data set. Our evaluation approach relies on two building
blocks that we introduce for the TFHE scheme:

\begin{itemize}
  
  \item A homomorphic controlled demultiplexer gate (\textbf{CDEMUX}), that
    performs the inverse operation of TFHE's multiplexer gate (CMUX~\cite{chillotti:2020}) with
    similar operands.

  \item An \textbf{Inverse Vertical Packing} (IVP) technique, based on the
    controlled demultiplexer gate (CDEMUX) and other operations of TFHE, which
    produces or updates a LUT, from which one may later evaluate using TFHE's
    Vertical Packing (VP)~\cite{chillotti:2017} technique.

\end{itemize}

Our homomorphic evaluation of the training procedure excels not only for its
practical results in terms of performance and accuracy but also for its
versatility and the simplicity of its implementation. Particularly, we show it
can be easily employed in scenarios such as distributed,
federated~\cite{zhang:2021}, and continuous~\cite{parisi:2019} learning. We
also define additional procedures, such as an efficient method for
homomorphically performing dataset balancing. 


\subsubsection{Implementation and Results}

We benchmark our construction over the main datasets used in the literature,
and we show significant improvements in encrypted training performance with a
small impact on the accuracy. We provide several options for training
parameters that trade off accuracy and performance. For the MNIST dataset, our
solutions enable accuracy varying from 91.71\% up to 93.76\% with execution
time from just 3.5 minutes up to 3.5 hours. This represents a significant
performance improvement over previous literature (which would take from 1.5 to
8 days) at the cost of an accuracy drop of 2.5\% to 5.4\%. For the HAM10000
dataset~\cite{tschandl:2018}, we improve both performance and accuracy compared
to previous literature, achieving 67.85\% to 69.85\% accuracy with encrypted
training time varying from 1.5 minutes up to 1 hour. Compared to
Glyph~\cite{lou:2020}, this represents an accuracy improvement of 0.65\% with a
performance improvement of 60 times (already adjusting for differences in
the execution environment).

We provide an open-source proof-of-concept implementation of training and
evaluation procedures \anon{based on the MOSFHET~\cite{guimaraes:2022b}
library, available at \url{https://github.com/leonardohn/homomorphic-wisards}}.


\subsubsection{Further Improvements}

We see end-to-end encrypted training as the major challenge for enabling
practical privacy-preserving machine learning. Therefore, this work focuses
strictly on this problem, without further exploring techniques that would
improve accuracy or performance but also require additional assumptions
(\emph{e.g.}, transfer learning, client-assisted training, and multi-party
protocols). Nonetheless, it is important to notice that these techniques are
not alternatives to our proposals. On the contrary, they are context-specific
optimizations that could be applied over WNNs in similar ways as they are for
general neural networks. Transfer learning, for example, which has been shown
to bring significant improvements for the encrypted training of
CNNs~\cite{lou:2020,lee:2023,panzade:2024}, could also be straightforwardly
applied to WNNs~\cite{milhomem:2018} for problems in which it fits. The same
can be said for client-assisted computation, which can generally be used to
accelerate most HE workloads.


\paragraph{Paper Structure}

The rest of the paper is organized as follows. Section 2 describes the relevant theoretical basis. Section 3 introduces the Homomorphic WiSARD architecture. Section 4 presents our results, comparing with the current state of the art. Section~\ref{sec:conclusion} concludes the paper.


\section{Background}


\subsection{Notation}

Let $\mathbb{Z}$ be the set of integer numbers and $\mathcal{R}$ be a
polynomial ring $\mathcal{R} = \mathbb{Z}[X]/(X^N + 1)$ with coefficients in
$\mathbb{Z}$ modulo some power-of-two cyclotomic polynomial $(X^N + 1)$. We use
subscript to denote their moduli and superscript to denote the number of
dimensions, for example $\mathbb{Z}_q^n$ is the set of vectors of size $n$
with elements in $\mathbb{Z}/q\mathbb{Z}$ (the integers modulo $q$). For
clarity, we use brackets to index neural network models represented as
multidimensional matrices, and subscript to index everything else (\emph{e.g.},
list, vectors, and tuples). We denote Uniform and Gaussian sampling with mean
zero and standard deviation $\sigma$ from some group $\mathbb{G}$ respectively
as $x \stackrel{\mathrm{U}}{\longleftarrow} \mathbb{G}$ and
$x \stackrel{\mathrm{G(\sigma)}}{\longleftarrow} \mathbb{G}$.


\subsection{Weightless Neural Networks}

Weightless Neural Networks (WNNs) (also known as $N$-tuple classifiers~\cite{bledsoe:1959}
or \textit{RAMnets}) are one of the oldest neural network-like algorithms ever
created for image recognition. Training and inference are entirely based on
programming Random Access Memory units (RAMs), which are mutable $n$-bit input,
$2^n$-output lookup tables, that may be configured to behave as any discrete
function. 



By itself, a single RAM could already be considered a classification algorithm.
It can learn discrete functions by adjusting their outputs in response to
input-output pairs from a target function and evaluate any discrete function in
constant time. This expressiveness, however, comes with the downside that the
RAM size doubles for every bit we add to the input, quickly becoming infeasible
for large inputs. As there might be no representing samples for certain input
permutations in a training set, RAMs are also unable to generalize over unseen
inputs.


\subsection{WiSARD Model}

In this work, we start with one of the most basic instances of a WNN: the
Wilkie, Stonham, and Aleksander’s Recognition Device
(WiSARD)~\cite{aleksander:1984}. Created to address the limitations of
freestanding RAMs, the WiSARD model organizes multiple RAMs into structures
denoted discriminators. Each discriminator partitions the input into $k$
$n$-bit tuples, which are fed into $k$ distinct RAMs. The combined outputs from
the RAMs of a discriminator, denoted as the score, quantify the recognized
sub-patterns present in the input. In a classification problem, each
discriminator represents a class, and the class corresponding to the
discriminator with the highest score is taken as the prediction.
Definition~\ref{def:wisard} describes a generic version of the WiSARD model for classification problems. For the original model, $\mathbb{G}$ is the set of
binary numbers $\mathbb{B}$.

\begin{definition}[Generic WiSARD Model]%
  \label{def:wisard}
  
  Given a classification problem for input samples of bit size $(s)$ and $(l)$
  classes, a Generic WiSARD model $\mathcal{W}_{(s, l, a, r)}$ with address
  size $(a)$ and random seed $(r)$ is a matrix in $\mathbb{G}^{l\times k \times
  2^{a}}$ with elements in some group $\mathbb{G}$. It comprises $l$
  discriminators, such that each discriminator is a tuple of $k = \left\lceil s
  \mathbin{/} a \right\rceil$ RAMs, and each RAM is a tuple of $2^a$ elements
  in $\mathbb{G}$. Let $\pi_r: \mathbb{B}^s \mapsto \mathbb{B}^s$ be a
  pseudo-random bit permutation map deterministic on the value $r$, and
  $f_{\mathsf{comp}}: \mathbb{B}^s \times \mathbb{Z} \mapsto \mathbb{Z}_{2^a}$
  be a composition function given by $f_{\mathsf{comp}} (x, d) =
  \sum_{i=0}^{\textsc{Min}(a, s-ad)} x_{i+ad} 2^i$,
  Algorithms~\ref{alg:wis-train} and~\ref{alg:wis-eval} define the training and
  evaluation for $\mathcal{W}$, respectively. 

\end{definition}






\begin{algorithm}[ht]
  \caption{Training of a WiSARD model}
  \label{alg:wis-train}%
  \SetKwInOut{Input}{Input}
  \SetKwInOut{Output}{Output}
  \Input{a WiSARD model $\mathcal{W}_{(s, l, a, r)}$}
  \Input{a training set $T$ of size $n$ and its list of labels $L$}
  \Output{trained WiSARD model}
  \BlankLine
  $\mathcal{W} \gets \{0\};
  $ $r \stackrel{\mathrm{U}}{\gets} \mathbb{Z};
  $ $k \gets \lceil s / a \rceil$\\
  \For{$i \in [\![0, n)$}{
    $t \gets \pi_r(T_i) $ \\
    \For{$j \in [\![0, k)$}{
      $ d \gets f_{\mathsf{comp}}(t, j) $ \\
      $\mathcal{W}[L_i][j][d] \gets 1$ \\
    }
  }
   \Return $\mathcal{W}$
\end{algorithm}



\begin{algorithm}[ht]
  \caption{Evaluate a sample in a WiSARD model}
  \label{alg:wis-eval}%
  \SetKwInOut{Input}{Input}
  \SetKwInOut{Output}{Output}
  \Input{a WiSARD Model $\mathcal{W}_{(s, l, a, r)}$}
  \Input{a sample $t$}
  \Output{classification of $t$}
  \BlankLine
  $k \gets \lceil s / a \rceil$;
  $t'\gets \pi_r(t)$ \\
  \For{$j \in l$}{
    $\displaystyle u_j \gets \sum_{i = 0}^{k} \mathcal{W}[j][i][f_{\mathsf{comp}}(t', i)]$\\
  }
  \Return \textsc{ArgMax}(u)
\end{algorithm}

\begin{figure}[ht]
  \centering
  \begin{tikzpicture}

  \definecolor{lppink}{HTML}{f4e0e9}
  \definecolor{lpblue}{HTML}{cfdaf0}
  \definecolor{lpgreen}{HTML}{d7e0b1}
  \definecolor{lporange}{HTML}{f4e3c9}

  \tikzstyle{myrect} = [
    draw,
    rectangle split,
    rectangle split parts=#1,
    rectangle split part align=left,
    text width=1em,
    align=center,
  ]

  \foreach \z in {.9em, .6em, .3em} {
    \node [
      myrect=4,
      fill=lporange,
      xshift=\z,
      yshift=\z,
    ] (temp) {
      \strut
      \nodepart{two}\strut
      \nodepart{three}\strut
      \nodepart{four}\strut
    };

    \node [
      myrect=4,
      fill=lporange,
      below=0.2em of temp,
    ] {
      \strut
      \nodepart{two}\strut
      \nodepart{three}\strut
      \nodepart{four}\strut
    };
  };

  \node [
    myrect=4,
    fill=lporange,
    label={[label distance=1.2em, xshift=.5em]\footnotesize RAMs},
  ] (ram1) {
    \strut 0
    \nodepart{two}\strut 1
    \nodepart{three}\strut 1
    \nodepart{four}\strut 0
  };

  \node[anchor=east] at (ram1.text west) {00};
  \node[anchor=east, draw] (dest1) at (ram1.two west) {01};
  \node[anchor=east] at (ram1.three west) {10};
  \node[anchor=east] at (ram1.four west) {11};

  \node [
    myrect=4,
    fill=lporange,
    below=0.2em of ram1,
    label={[label distance=0.5em, xshift=1em]270:\footnotesize Discriminators},
  ] (ram2) {
    \strut 0
    \nodepart{two}\strut 0
    \nodepart{three}\strut 0
    \nodepart{four}\strut 1
  };

  \node[anchor=east] at (ram2.text west) {00};
  \node[anchor=east] at (ram2.two west) {01};
  \node[anchor=east, draw] (dest2) at (ram2.three west) {10};
  \node[anchor=east] at (ram2.four west) {11};

  \coordinate(midram) at ($(ram1)!0.5!(ram2)$);

  \draw[dashed] ($(midram)-(3.7em,8.2em)$)
    rectangle ($(midram)+(5.5em,10.2em)$);

  \draw[dashed] ($(midram)-(11.3em,4.2em)$)
    rectangle ($(midram)+(-4em,5.8em)$);

  \node [
    myrect=4,
    fill=lpgreen,
    text width=1em,
    align=center,
    left=5em of midram,
    label={[label distance=0.5em]\footnotesize Address},
  ] (address) {
    \strut 0
    \nodepart{two}\strut 1
    \nodepart{three}\strut 1
    \nodepart{four}\strut 0
  };

  \node [
    myrect=4,
    fill=lpblue,
    text width=1em,
    align=center,
    left=2.3em of address,
    label={[label distance=.5em]\footnotesize Input},
    label={[label distance=.5em, xshift=2.2em]270:\footnotesize Permutation},
  ] (input) {
    \strut 1
    \nodepart{two}\strut 0
    \nodepart{three}\strut 1
    \nodepart{four}\strut 0
  };

  \draw[-latex] (input.text east) -- +(.3em,0)
    -- ($(address.three west)-(.7em,0)$) -- (address.three west);
  \draw[-latex] (input.two east) -- +(.3em,0)
    -- ($(address.text west)-(.7em,0)$) -- (address.text west);
  \draw[-latex] (input.three east) -- +(.3em,0)
    -- ($(address.two west)-(.7em,0)$) -- (address.two west);
  \draw[-latex] (input.four east) -- +(.3em,0)
    -- ($(address.four west)-(.7em,0)$) -- (address.four west);

  \draw (address.text east) -- +(1.9em,0);
  \draw[-latex] (address.two east) -- +(1.9em,0) |- (dest1.west);
  \draw[-latex] (address.three east) -- +(1.9em,0) |- (dest2.west);
  \draw (address.four east) -- +(1.9em,0);

  \foreach \z in {.9em, .6em, .3em} {
    \node [
      draw,
      circle,
      fill=lppink,
      align=center,
      right=2.2em of midram,
      xshift=0.6667*\z,
      yshift=0.6667*\z,
    ] (temp) {$\sum$};
    \draw[-latex] ($(ram1.two east)+(\z,\z)$) -- +(.7em,0) -- (temp);
    \draw[-latex] ($(ram2.three east)+(\z,\z)$) -- +(.7em,0) -- (temp);
    \draw[-latex] (temp.east) -- +(1.75em-0.6667*\z,0);
  };

  \node [
    draw,
    circle,
    fill=lppink,
    align=center,
    right=2.2em of midram,
  ] (disc) {$\sum$};

  \draw[-latex] (ram1.two east) -- +(.7em,0) -- (disc);
  \draw[-latex] (ram2.three east) -- +(.7em,0) -- (disc);
  \draw[-latex] (disc.east) -- +(1.75em,0);

  \coordinate (plot) at ($(midram)+(6.6em,-3em)$);

  \begin{axis} [
    width=15em,
    height=10.6em,
    xbar=.05cm,
    bar width=1em,
    xmin=0,
    xmax=6,
    ticks=none,
    enlarge x limits={value=.25, upper},
    enlarge y limits={abs=.8},
    at=(plot),
    legend style={font=\scriptsize},
    xlabel={\footnotesize Scoreboard},
    x label style={at={(axis description cs:0.5,-0.05)},anchor=north},
  ]

    \addplot[fill=lppink] coordinates {(3.8,1)};
    \addplot[fill=lpblue] coordinates {(1.8,1)};
    \addplot[fill=lpgreen] coordinates {(2.8,1)};
    \addplot[fill=lporange] coordinates {(0.8,1)};
    
    \legend {Class 1, Class 2, Class 3, Class 4};

  \end{axis}

\end{tikzpicture}
  \caption{Evaluation of a WiSARD model.}
  \label{fig:wisard}
\end{figure}
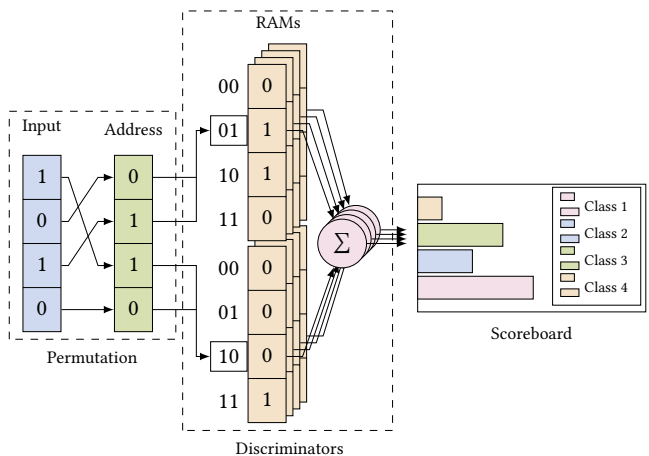

The basic idea behind the inference using the WiSARD model is that each RAM
would recognize small bit patterns to classify the input. Notice that, during
training, each discriminator receives only samples from its respective label,
learning to identify these small bit patterns. During the evaluation phase, the
input is presented to all discriminators in the model, each producing a score
based on the values provided by their RAMs. 
Figures~\ref{fig:wisard}~and~\ref{fig:discr} depict the entire inference
process and examples of a discriminator with input size $s=4$ and address
size $a=2$, respectively.


\subsection{State-of-the-art WNNs}

The WiSARD model was first introduced in 1984 as a commercial version of the
$N$-tuple classification algorithm from 1959~\cite{bledsoe:1959}. Since then,
many techniques have been introduced to improve the WiSARD model and, more
generally, Weightless Neural Network algorithms. This section discusses some of
them.


\subsubsection{Bleaching}

Bleaching~\cite{grieco:2010} is a technique used in the WiSARD model to
adjust the sensitiveness of the model to repeated sub-patterns in the training
set. It consists of replacing the RAMs binary values with integers, which now
counts how many times small bit patterns are seen during the training phase.

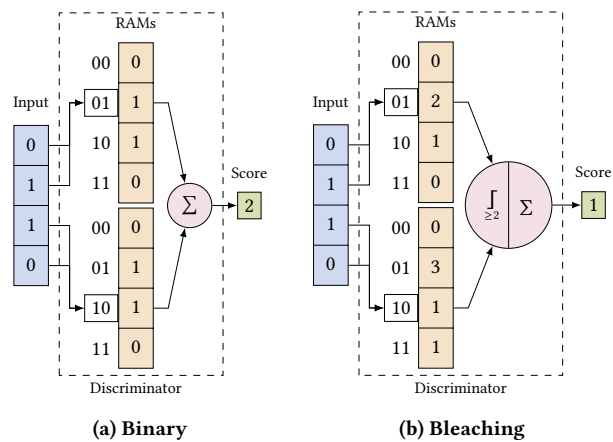
\begin{figure}[ht]
  \centering
  \begin{subfigure}{.42\linewidth}
    \centering
    \begin{tikzpicture}

  \definecolor{lppink}{HTML}{f4e0e9}
  \definecolor{lpblue}{HTML}{cfdaf0}
  \definecolor{lpgreen}{HTML}{d7e0b1}
  \definecolor{lporange}{HTML}{f4e3c9}

  \tikzstyle{myrect} = [
    draw,
    fill=lporange,
    rectangle split,
    rectangle split parts=#1,
    rectangle split part align=left,
  ]

  \node [
    myrect=4,
    text width=1em,
    align=center,
    label={[label distance=1mm]\footnotesize RAMs},
  ] (ram1) {
    \strut 0
    \nodepart{two}\strut 1
    \nodepart{three}\strut 1
    \nodepart{four}\strut 0
  };

  \node[anchor=east] at (ram1.text west) {00};
  \node[anchor=east, draw] (dest1) at (ram1.two west) {01};
  \node[anchor=east] at (ram1.three west) {10};
  \node[anchor=east] at (ram1.four west) {11};

  \node [
    myrect=4,
    below=0.2em of ram1,
    text width=1em,
    align=center,
    label={[label distance=1mm]below:\footnotesize Discriminator},
  ] (ram2) {
    \strut 0
    \nodepart{two}\strut 1
    \nodepart{three}\strut 1
    \nodepart{four}\strut 0
  };

  \node[anchor=east] at (ram2.text west) {00};
  \node[anchor=east] at (ram2.two west) {01};
  \node[anchor=east, draw] (dest2) at (ram2.three west) {10};
  \node[anchor=east] at (ram2.four west) {11};

  \coordinate(midram) at ($(ram1)!0.5!(ram2)$);

  \draw[dashed] ($(midram)-(3.7em,8.2em)$)
    rectangle ($(midram)+(4.2em,9.3em)$);

  \node [
    myrect=4,
    fill=lpblue,
    text width=1em,
    align=center,
    left=4.2em of midram,
    label={[label distance=1mm]\footnotesize Input},
  ] (address) {
    \strut 0
    \nodepart{two}\strut 1
    \nodepart{three}\strut 1
    \nodepart{four}\strut 0
  };

  \draw[-latex] (address.text east) -- +(1em,0) |- (dest1.west);
  \draw[-latex] (address.two east) -- +(1em,0) |- (dest1.west);

  \draw[-latex] (address.three east) -- +(1em,0) |- (dest2.west);
  \draw[-latex] (address.four east) -- +(1em,0) |- (dest2.west);

  \node [
    draw,
    circle,
    fill=lppink,
    align=center,
    right=1.5em of midram,
  ] (disc) {$\sum$};

  \draw[-latex] (ram1.two east) -- +(.7em,0) -- (disc);
  \draw[-latex] (ram2.three east) -- +(.7em,0) -- (disc);

  \node [
    draw,
    fill=lpgreen,
    align=center,
    right=1.2em of disc,
    label={[label distance=1mm]\footnotesize Score},
  ] (score) {2};

  \draw[-latex] (disc.east) -- (score.west);

\end{tikzpicture}
    \caption{Binary}
    \label{fig:discr}
  \end{subfigure}%
  \begin{subfigure}{.05\linewidth}
    \hfill
  \end{subfigure}%
  \begin{subfigure}{.49\linewidth}
    \centering
    \begin{tikzpicture}

  \definecolor{lppink}{HTML}{f4e0e9}
  \definecolor{lpblue}{HTML}{cfdaf0}
  \definecolor{lpgreen}{HTML}{d7e0b1}
  \definecolor{lporange}{HTML}{f4e3c9}

  \tikzstyle{myrect} = [
    draw,
    fill=lporange,
    rectangle split,
    rectangle split parts=#1,
    rectangle split part align=left,
  ]

  \node [
    myrect=4,
    text width=1em,
    align=center,
    label={[label distance=1mm]\footnotesize RAMs},
  ] (ram1) {
    \strut 0
    \nodepart{two}\strut 2
    \nodepart{three}\strut 1
    \nodepart{four}\strut 0
  };

  \node[anchor=east] at (ram1.text west) {00};
  \node[anchor=east, draw] (dest1) at (ram1.two west) {01};
  \node[anchor=east] at (ram1.three west) {10};
  \node[anchor=east] at (ram1.four west) {11};

  \node [
    myrect=4,
    below=0.2em of ram1,
    text width=1em,
    align=center,
    label={[label distance=1mm, xshift=1.2em]below:\footnotesize Discriminator},
  ] (ram2) {
    \strut 0
    \nodepart{two}\strut 3
    \nodepart{three}\strut 1
    \nodepart{four}\strut 1
  };

  \node[anchor=east] at (ram2.text west) {00};
  \node[anchor=east] at (ram2.two west) {01};
  \node[anchor=east, draw] (dest2) at (ram2.three west) {10};
  \node[anchor=east] at (ram2.four west) {11};

  \coordinate(midram) at ($(ram1)!0.5!(ram2)$);

  \draw[dashed] ($(midram)-(3.7em,8.2em)$)
    rectangle ($(midram)+(6.1em,9.3em)$);

  \node [
    myrect=4,
    fill=lpblue,
    text width=1em,
    align=center,
    left=4.2em of midram,
    label={[label distance=1mm]\footnotesize Input},
  ] (address) {
    \strut 0
    \nodepart{two}\strut 1
    \nodepart{three}\strut 1
    \nodepart{four}\strut 0
  };

  \draw[-latex] (address.text east) -- +(1em,0) |- (dest1.west);
  \draw[-latex] (address.two east) -- +(1em,0) |- (dest1.west);

  \draw[-latex] (address.three east) -- +(1em,0) |- (dest2.west);
  \draw[-latex] (address.four east) -- +(1em,0) |- (dest2.west);

  \coordinate(sum) at ($(midram)+(3.5em,0)$);

  \node [
    draw,
    circle split,
    rotate=90,
    fill=lppink,
    align=center,
    inner sep=0.5em,
  ] (disc) at (sum) {
    \rotatebox{-90}{$\underaccent{\geq 2}{\rfloor\!\!\lceil}$}
    \nodepart{lower}%
    \rotatebox{-90}{\large$\Sigma$}
  };

  \draw[-latex] (ram1.two east) -- +(.7em,0) -- (disc);
  \draw[-latex] (ram2.three east) -- +(.7em,0) -- (disc);

  \node [
    draw,
    fill=lpgreen,
    align=center,
    right=3.5em of sum,
    label={[label distance=1mm]\footnotesize Score},
  ] (score) {1};

  \draw[-latex] (disc.south) -- (score.west);

\end{tikzpicture}
    \caption{Bleaching}
    \label{fig:bleaching}
  \end{subfigure}%
  \caption{Evaluation of two discriminators.}
\end{figure}

During the evaluation phase, a threshold operator is applied to every RAM
output value, resulting in a binary value that will be used to produce the
score. This adjustment can significantly improve the accuracy of models with
smaller address sizes. In Figure~\ref{fig:bleaching} we illustrate the
evaluation of a discriminator with bleaching.

\subsubsection{Quantization}

Quantization~\cite{keutzer:2022} is a (possibly lossy) transformation constraining input values from a
continuous or otherwise large set of values to a discrete, typically smaller
set. Examples of quantization functions are as follows.

\begin{itemize}
  \item\textbf{Linear}: $f_r(x) = \lfloor x / r \rfloor$,
    for a constant ratio $r$.
  \item\textbf{Logarithmic}: $f_b(x) = \lfloor \log_b(x + 1) \rfloor$,
    for a base $b$.
\end{itemize}

The simplest form is linear quantization, where the input values are mapped to
uniformly distributed values along the output domain. Other methods include
logarithmic and Gaussian quantization, which can be beneficial when the sample
density distribution is known. Choosing a distribution that better
distinguishes values along key regions in the value space may help minimize the
inherent rounding error introduced by this transformation.


\subsubsection{Thermometer Encoding}

Thermometer encoding~\cite{buckman:2018}, sometimes referred to as unary encoding, represents
natural numbers, starting from zero, as increasing sequences of bits with value
one. A thermometer $\mathcal{T}_N: \mathbb{N} \mapsto \mathbb{B}^N$ encodes a
value $x \in \mathbb{N}$ into a vector $t = \mathcal{T}_N(x)$ such that $t_i =
1$ for $i \in [0, x)$, and $t_i = 0$ for $i \in [x, N)$. For instance,
$\mathcal{T}_4(3) = [1, 1, 1, 0]$ and $\mathcal{T}_4(1) = [1, 0, 0, 0]$. The
name comes from the resemblance to a thermometer, which fills in response to a
temperature increase.


\subsection{Homomorphic Encryption}

Homomorphic encryption (HE) is a technique that allows computation over
encrypted data by establishing a map (homomorphism) between operations over the
ciphertexts and the messages they encrypt. It provides similar confidentiality
guarantees as traditional encryption schemes, fully protecting data during
computation. Most of the modern HE schemes are based on the Learning With
Errors (LWE)~\cite{regev:2009} problem and its ring variant
(RLWE)~\cite{lyubashevsky:2010}.


\subsubsection{TFHE Scheme}%
\label{sec:tfhedef}

We opt to work with the TFHE~\cite{chillotti:2020} scheme, as it presents
techniques for efficiently evaluating lookup tables which are a perfect match
for WiSARD's RAM-based logic. Furthermore, as we will show in
Section~\ref{sec:homwis}, it can also be used to efficiently generate RAM units
during training. In this work, we only use a subset of the operations the
scheme provides, which we define in the following.

\begin{itemize}
  
  \item $\textsc{Setup}(\lambda)$: Security parameters ($q, N, n, \sigma_{0},
    \sigma_{1}$) are defined based on the plaintext space $\mathbb{Z}_p$, on
    the circuit to be homomorphically evaluated, and on security level, which
    we estimate using the Lattice Estimator~\cite{albrecht:2015}. The
    polynomial ring $\mathcal{R}$ is defined as $\mathcal{R} =
    \mathbb{Z}[X]/(X^N + 1)$.
  
  \item $\textsc{SecretKeyGen}(N, n)$: Given the lattices dimensions $n$ and
    $N$, it samples an LWE key $S_{0} \stackrel{\mathrm{U}}{\gets}
    \mathbb{B}^n$ and an RLWE key $S_{1} \stackrel{\mathrm{U}}{\gets}
    \mathcal{R}_2$ and returns a pair of keys $(S_{0}, S_{1})$.
  
  \item Encryption:
    \begin{itemize}
      
      \item $\textsc{EncLWE}(m, S_0, \sigma_0)$: Given the LWE secret key
        $S_0$, a message $m \in \mathbb{Z}_p$, and noise parameter $\sigma_0$,
        it samples $a \stackrel{\mathrm{U}}{\gets} \mathbb{Z}_q^n$ and $e
        \stackrel{\mathrm{G(\sigma_0)}}{\longleftarrow} \mathbb{Z}_q$, and
        computes $b \gets \langle a, S_0 \rangle + m\frac{q}{p} + e$ to produce
        the LWE ciphertext $(a,b) \in \text{LWE}_{S_0}(m)$.
      
      \item $\textsc{EncRLWE}(m, S_1, \sigma_1)$: Given the RLWE secret key
        $S_1$, a message $m \in \mathcal{R}_p$, and noise parameter $\sigma_1$,
        it samples $a \stackrel{\mathrm{U}}{\gets} \mathcal{R}_q$ and $e
        \stackrel{\mathrm{G(\sigma_1)}}{\longleftarrow} \mathcal{R}_q$, and
        computes $b \gets a \cdot S_1 + m\frac{q}{p} + e$ to produce the RLWE
        ciphertext $(a,b) \in \text{RLWE}_{S_1}(m)$.
      
      \item $\textsc{EncRGSW}(m, S_1, \sigma_1)$: Given the RLWE secret key
        $S_1$, a message $m \in \mathcal{R}_p$, a noise parameter $\sigma_1$,
        and decomposition parameters $\ell$ and $\beta$, it produces a vector
        $C \in \mathcal{R}_q^{2\ell \times 2}$ of $2\ell$ RLWE ciphertexts
        encrypting 0's and a gadget decomposition matrix $G = [[0,
        \frac{q}{p\beta^0}], \dots, [0, \frac{q}{p\beta^{\ell-1}}],\allowbreak
        [\frac{q}{p\beta^0}, 0], \dots, [\frac{q}{p\beta^{\ell-1}}, 0]] \in
        \mathcal{R}_q^{2\ell \times 2}$. It returns the RGSW ciphertext given
        by $(C + m \cdot G) \in \text{RGSW}_{S_1}(m)$.
    
    \end{itemize}
  
  \item Decryption:
    \begin{itemize}
      
      \item $\textsc{DecLWE}(c = (a,b), S_0)$: Returns $\left\lceil \frac{b -
        \langle a, S_0 \rangle}{q/p} \right\rfloor$
      
      \item $\textsc{DecRLWE}(c = (a,b), S_1)$: Returns $\left\lceil \frac{b -
        a \cdot S_1}{q/p} \right\rfloor$
      
      \item $\textsc{DecRGSW}(c = (c_0, c_1, \dots, c_{2\ell - 1}), S_1)$:\\
        Returns $\textsc{DecRLWE}(c_0, S_1)$
    
    \end{itemize}

  \item Evaluation:
    \begin{itemize}
      
      \item Addition ($c_0 + c_1$): Given ciphertexts $c_0$ and $c_1$ of the
        same type encrypting messages $m_0$ and $m_1$, it returns a ciphertext
        encrypting $m_0 + m_1$.
      
      \item Muliplication ($C_0 \boxdot c_1$): Given an RGSW ciphertext $C_0$
        and an RLWE ciphertext $c_1$ encrypting messages $m_0$ and $m_1$,
        respectively, it returns an RLWE ciphertext encrypting $m_0 \cdot m_1$.

       \item CMUX($C_0, c_1, c_2$): Given an RGSW ciphertext $C_0$
        and RLWE ciphertexts $c_1$ and $c_2$ encrypting messages $m_0 \in \{0,1\}$, $m_1 \in \mathcal{R}_p$ and $m_2 \in \mathcal{R}_p$, respectively, it returns an RLWE ciphertext encrypting $m_0 \cdot m_1 + (1 - m_0) \cdot m_2$, which represents the computation of a selection (multiplexer) between $m_1$ and $m_2$ with selector $m_0$.
      
      \item ExtractLWE($c_0$, $i$): Given an integer $i$ and an RLWE ciphertext
        $c_0$ encrypting a polynomial $v = \sum_{j=0}^{N-1} m_jX^j$ under key
        $S$, it returns an LWE ciphertext encrypting $m_i$.
      
      \item PackingKeySwitching($c$): Given a vector of $k$ LWE ciphertexts $c$
        with each of them encrypting a message $m_i \in \mathbb{Z}_p$, returns
        an RLWE ciphertext encrypting the polynomial $\sum^k_{i=0} m_i X^i \in
        \mathcal{R}_p$.
      
      \item BlindRotate($c$, $C$, $I$): Given a vector of $k$ RGSW ciphertexts
        $C$ with each of them encrypting a binary message $S_i \in \mathbb{B}$,
        a vector of k integers $I$, a RLWE ciphertext $c$ encrypting a
        polynomial $v$, the BlindRotate returns an RLWE ciphertext $c'\in
        \text{RLWE}\left(vX^{-\sum_{i=0}^{k-1}{S_iI_i}}\right)$, which is
        negacyclic rotation of $v$.
    
    \end{itemize}

\end{itemize}

We only define the operations we use in this paper, and we treat them as
black-box procedures all those in which the inner details do not affect our
proposals. For more details about the TFHE scheme, refer
to~\cite{chillotti:2017} and~\cite{chillotti:2020}. In the HE literature,
ciphertexts are commonly referred to as \textit{``samples''}, a term that is
also used in the ML literature to refer to data objects that are input to a
machine learning algorithm. To avoid confusion, we use the term
\textit{``sample''} only for the input data of an ML algorithm and not for
ciphertexts.


\subsubsection{LUT Evaluation}

The main feature provided by TFHE is the efficient homomorphic evaluation of
lookup tables (LUTs), which can represent any discretized function. TFHE
presents two main methods for LUT evaluation, which we describe in the
following.

\paragraph{Vertical Packing (VP)}~\cite{chillotti:2017}: Let $C$ be a list of RGSW
ciphertext encrypting bit by bit the input $g \in \mathbb{B}^s$ to some
function $f: \mathbb{B}^s \mapsto \mathbb{Z}_p$, the vertical evaluates $f$ as
follows:

\begin{enumerate}
  
  \item Function encoding: Let $v\in \mathbb{Z}_p^{2^s}$ be a vector of
    evaluations of $f$ such that $v_i = f(i)$ for all $i \in \mathbb{B}^s$, it
    encrypts $v$ in a vector of $\lceil s/N \rceil$ RLWE ciphertexts $L$, such
    that each RLWE ciphertext $L_i$ encrypts the polynomial $\sum_{j=0}^{N}
    v_{iN + j} X^j$.
  
  \item CMUX tree: Let $\overline{C}$ be a vector given by the first
    $\lceil\log_2(s/N)\rceil$ elements of $C$, and given the encrypted LUT $L$,
    the CMUX tree is defined in lines 2 to 6 of Algorithm~\ref{alg:vp}. The
    vertical packing uses the CMUX tree to select the RLWE sample containing
    the desired position, $L_0$. 
  
  \item \textsc{BlindRotate} and result extraction: Let $\underline{C}$ be the
    vector given by the last $\log_2(N)$ elements of $C$, and given the RLWE
    sample $c$ encrypting the polynomial $v$ containing the desired position,
    the VP uses \textsc{BlindRotate}$(c, \underline{C}, [2^0, 2^1, \allowbreak
    2^2, \dots, 2^{\log_2(N) - 1}])$ to rotate the desired position to the
    constant term of $v$, obtaining an RLWE sample $\hat{L}$. It then performs
    \textsc{ExtractLWE}($\hat{L}$, 0) to extract the desired position to an LWE
    ciphertext, which is the computation result.

\end{enumerate}

\begin{algorithm}[ht]
  
  \caption{Vertical Packing}
  \label{alg:vp}%
  \SetKwInOut{Input}{Input}
  \SetKwInOut{Output}{Output}
  
  \Input{
    a list of $k$ RGSW ciphertexts $C$, such that each $C_i$ encrypts a message
    $m_i \in \mathbb{B}$ for $i \in [\![0,k)$, where $m_i$ is the bit
    decomposition of some message $m = \sum_{i=0}^{k-1} m_i2^i$
  }
  \Input{
    a list of $2^z$ RLWE ciphertexts $L$, each with RLWE dimension $N$,
    encrypting a lookup table containing evaluations of some function $f :
    \mathbb{Z}_{2^k} \mapsto \mathbb{Z}_p$
  }
  \Output{an LWE sample encrypting $f(m)$}
  
  \BlankLine
  $n \gets 2^z/2$\\
  \For{$i \gets 0$ \textbf{ to } $z - 1$}{
    \For{$j \gets 0$ \textbf{ to } $n$}{
    $L_j \gets \textsc{CMUX}(C_i, L_{j}, L_{j + n})$ \\
    }
    $n \gets n/2$
  }
  $\hat{L} \gets $\textsc{BlindRotate}$
    (L_0, [C_z, \dots, C_k],  [2^0, 2^1, \dots, N/4, N/2])$ \\
  \Return \textsc{ExtractLWE}$(\hat{L}, 0)$

\end{algorithm}


\subsubsection{Programmable Bootstrapping}

As described in Section~\ref{sec:tfhedef}, LWE-based encryption requires the
addition of a noise (error) component to provide security. This noise grows
with every arithmetic operation and eventually needs to reset to some small
value to allow for new operations. This process of resetting the noise is a
\textit{bootstrapping}. In schemes such as TFHE~\cite{chillotti:2020} and
FHEW~\cite{ducas:2015}, the bootstrapping is implemented as LUT evaluation,
which allows one to also use them to evaluate arbitrary functions, a process
called \textit{functional boostrapping}~\cite{boura:2019}, or, more
specifically for TFHE, \textit{programmable
boostrapping}~\cite{chillotti:2021}. For this work, we see it as just another
way of evaluating arbitrary functions represented as LUTs using TFHE. Different
from the VP, it does not require RGSW ciphertexts, being more flexible and
allowing its use in composed circuits. On the other hand, it is significantly
more expensive and only capable of evaluating small LUTs with good
performance.

The evaluation of arbitrary functions using the programmable bootstrapping is
broadly explored in the literature~\cite{guimaraes:2021, chillotti:2021}.
Therefore, we take it as a black-box procedure for the homomorphic evaluation
of the following functions:

\begin{itemize}
  
  \item \textsc{EncryptedArgMax}: Given an array of LWE ciphertexts, it returns
    an LWE ciphertext encrypting the index of the sample with the highest
    value.
  
  \item \textsc{Activate}: Given a function $f$ and an array of LWE ciphertexts
    encrypting messages $m_i$, returns an array of LWE ciphertexts encrypting
    messages $f(m_i)$.

\end{itemize}


\section{Homomorphic WiSARDs}%
\label{sec:homwis}

The core procedure in WNNs is the evaluation of RAM units, which can be
efficiently evaluated by using some of TFHE's techniques for LUT evaluation.
However, training in WNNs still mixes arithmetic and non-arithmetic operations.
Considering this, our first step for homomorphically evaluating WNN is to
separate these operations into two independent procedures.


\subsection{Integer WiSARDs}

Let $\mathcal{W}_{(s, l, a, r)}$ be a WiSARD model as in
Definition~\ref{def:wisard}, the Integer WiSARD model follows the same
definition with $\mathbb{G} = \mathbb{Z}$ and the training process described in
Algorithm~\ref{alg:Iwistrain}. Notice that everything remains essentially the
same except line~\ref{line:add}, which now uses integer RAMs to count
occurrences of each input pattern. With this change, the procedure only 
requires linear arithmetic on each RAM element.

\begin{algorithm}[ht]
  
  \caption{Integer WiSARD model training}
  \label{alg:Iwistrain}%
  \SetKwInOut{Input}{Input}
  \SetKwInOut{Output}{Output}
  \Input{a WiSARD model $\mathcal{W}_{(s, l, a, r)}$}
  \Input{a training set $T$ of size $n$ and its list of labels $L$}
  \Output{trained Integer WiSARD model}
  
  \BlankLine
  $\mathcal{W} \gets \{0\};
  $ $r \stackrel{\mathrm{U}}{\gets} \mathbb{Z};
  $ $k \gets \lceil s / a \rceil$\;
  \For{$i \in [\![0, n)$}{
    $t \gets \pi_r(T_i) $ \\
    \For{$j \in [\![0, k)$}{
      $ d \gets f_{\mathsf{comp}}(t, j) $ \\
      $\mathcal{W}[L_i][j][d] \gets \mathcal{W}[L_i][j][d] + 1$ \\
      \label{line:add}
    }
  }
  \Return $\mathcal{W}$
\end{algorithm}

The inference algorithm remains the same as Algorithm~\ref{alg:wis-eval}. By
itself, this training process does not provide good accuracy levels. Once the
training is finished, we move to a Model Activation step, in which an
activation function is applied to each element of each RAM in $\mathcal{W}$.
Notice that together the integer training and RAM activation are functionally
equivalent to other WNNs. For example, we can obtain the original wizard by
taking a binarization function $f_{\mathsf{bin}} : \mathbb{Z} \mapsto
\mathbb{B}$, such that $f_{\mathsf{bin}}(x) = 1$ if $x > 0$, and $0$ otherwise,
as activation. In principle, we propose the model activation as a
post-processing to the training, but one could leave it for the inference
procedure, which may be more adequate depending on the scenario. It is
important to note that, different from typical NNs, the activation process
occurs over the model and not over the input data. It is a post-processing
procedure, similar, for example, to NN post-training quantization
processes~\cite{legiest:2023}, which also applies non-linear functions over the
NN model.


\subsection{Activated WiSARDs}

The literature on WNNs often focuses on computationally inexpensive functions
to program RAMs, as the goal of employing WNNs is generally to minimize the use
of computation resources during training. Binarization and threshold are the
main examples of those, and we consider both in our model activation step as a
way of evaluating the different existing WNN models. Nonetheless, once we treat
the activation as a separate procedure, we can also explore other functions and
methods for ``activating'' the model, obtaining an \textit{Activated WiSARD}.
Particularly, our HE evaluation is based on TFHE's LUT evaluation methods, for
which performance depends only on the function precision but not on the
specific functions being evaluated. 

Considering this, we experiment with more complex activation functions during
the model activation procedure. Our main result in this experiment is the
introduction of logarithmic activation for WiSARDs. It achieves superior
accuracy on problems such as digit recognition on the MNIST dataset compared to
traditional binary WiSARD models when using a large set of training samples.
Compared to threshold WiSARD models, it incurs a slight accuracy loss, with the
advantage of eliminating the need for threshold optimization, a process that
may require repeated training, which could introduce a significant slowdown for
the encrypted training. Our logarithmic (log) activation is defined as simply
applying the $f(x) = \log_2(x + 1)$ function over each integer element of the
RAMs of an Integer WiSARD model, whereas the bounded logarithmic (b-log)
activation uses the $f(x) = min(\log_2(x + 1), c)$, which is the $\log_2$
function with an upper limit value of some (typically small) constant $c > 0$.

\subsection{TFHE Building Blocks}

The process of evaluating a WiSARD model involves two primary steps: first, evaluating the RAMs, and then, aggregating their outputs to calculate the overall score. Evaluating a sample requires operations that can be easily implemented with standard LUT evaluation procedures from TFHE; Training a sample, on the other hand, requires dynamic LUT modifications to account for the sample. To achieve that, we define a functional inverse for the TFHE's Vertical Packing procedure, which we introduce in this section.


\subsubsection{CDEMUX Gate}

The Controlled Demultiplexer gate (CDEMUX) is the functional inverse of TFHE's
Controlled Multiplexer gate (CMUX). It is characterized by two input channels
and two output channels: a control input, that receives an RGSW ciphertext
encrypting a message in $\mathbb{B} = \{0, 1\}$, and a data input, which
receives an RLWE ciphertext. The two data output channels produce RLWE
ciphertexts. The CDEMUX gate, through homomorphic computation, directs the
input message to either the first or second output channel based on the value
of the control input message. Simultaneously, the alternate channel is defined
to be an RLWE ciphertext encrypting zero.

For TFHE, the implementation of the CDEMUX gate hinges on the utilization of
RGSW-RLWE external products. Mathematically, we formalize the $\text{CDEMUX
gate} : \text{RGSW} \times \text{RLWE} \mapsto \text{RLWE} \times \text{RLWE}$
as $\textsc{CDEMUX}(C, d_{in}) := (d_{in} -  C \boxdot d_{in}, C \boxdot
d_{in})$.


\subsubsection{CDEMUX Tree}

We may extend the CDEMUX gate to incorporate an arbitrary number of output
gates by employing a hierarchical, tree-like structure of CDEMUX gates, similar
to the CMUX tree in the Vertical Packing. The process initiates with a single
CDEMUX operation, followed by successive applications of CDEMUX to both outputs
recursively, each tree level utilizing a distinct control message. This
approach results in an array composed of RLWE ciphertexts, that is
predominantly encrypting zeros, except for one specific coefficient of a
specific ciphertext, defined with the value from the input data channel. The
position is determined by the values of the control messages used in each
level. This technique allows creating an array of any desired size, with
precise control over the initialization of a single secret position.


\subsubsection{Inverse Vertical Packing}

The Inverse Vertical Packing (IVP) technique uses the blind rotation operation
and the CDEMUX tree to create an encrypted single-valued LUT following TFHE's 
vertical packing encoding. Single-valued LUTs are those composed of zeros
except for one secretly designated cell. Complex LUTs can then be created by
superimposing multiple single-valued LUTs through RLWE summation.
Algorithm~\ref{alg:ivp} details the IVP procedure.

\begin{algorithm}[ht]
  
  \caption{Inverse Vertical Packing}
  \label{alg:ivp}%
  \SetKwInOut{Input}{Input}
  \SetKwInOut{Output}{Output}
  \SetKwFunction{proc}{CDEMUX}
  \Input{
    a list of $k$ RGSW ciphertexts $C$, such that each $C_i$ encrypts a message
    $m_i \mathbb{B}$ for $i \in [\![0,k)$, where $m_i$ is the bit decomposition
    of some message $m = \sum_{i=0}^{k-1} m_i2^i$
  }
  \Input{
    an RLWE sample $\hat{L}$ encrypting $f(m)$ for some function $f :
    \mathbb{Z}_{2^k} \mapsto \mathbb{Z}_p$
  }
  \Output{
    a list of $2^z$ RLWE ciphertexts $L$, each with RLWE dimension $N$,
    encrypting a lookup table containing 0 in all elements except the $m$-th,
    which encrypts $f(m)$
  }
  
  \BlankLine
  \setcounter{AlgoLine}{0}
  
  $L_0 \gets $\textsc{BlindRotate}$
    (\hat{L}, [C_z, \dots, C_k],  [-2^0, -2^1, \dots, -N/4, -N/2])$ \\
  $n \gets 1$ \\
  \For{$i \gets 0$ \KwTo $z - 1$}{
    \For{$j \gets 0$ \KwTo $n$}{
      $L_j, L_{j + n} \gets$ \proc{$C_i, L_j$}
    }
    $n \gets 2n$
  }
  \Return $L$ \\
  \BlankLine
  \SetKwProg{myproc}{Procedure}{}{}
    \setcounter{AlgoLine}{0}\myproc{\proc{$C_0,c_1$}}{
      $\hat{c} = C_0 \boxdot c_1$ \\
      \KwRet $(c_1 - \hat{c}, \hat{c})$ \\
    }

\end{algorithm}

Notice that, since the IVP receives an encrypted RLWE sample as input, it can
also be used to arbitrarily update an existing LUT. More specifically, one
could use the VP to select an element from the LUT, extract it to an LWE
sample, evaluate arbitrary functions over it, and, finally, use the IVP to move
it back to its original position.

\subsection{Homomorphic Training}

On the client side, we start with a pre-processing phase, where we perform
quantization and thermometer encoding over the input data. For each sample, we
only consider pre-processing techniques that are independent of the entire
dataset. For example, we avoid techniques such as sample average normalization,
which would require knowing data from other samples (which is not always
possible in scenarios where data is already encrypted or in cases such as
distributed, federated, or continuous learning~\cite{zhang:2021,parisi:2019}).
Each (quantized) input sample and its label are encrypted bit by bit as RGSW
ciphertexts. All data is sent to the server. Definition~\ref{def:homwisard}
formalizes the Homomorphic WiSARD Model. Compared to the standard model
(Definition~\ref{def:wisard}), the matrix representing this model is rearranged
(for compatibility with the vertical packing encoding) and encrypted in RLWE
ciphertexts. 

\begin{definition}[Homomorphic WiSARD Model]%
  \label{def:homwisard}

  Given a classification problem for encrypted input samples of bit size $s$
  and $l$ classes, a Homomorphic WiSARD model $\mathcal{H}_{(s, l, a, r)}$ with
  address size $a$ and random seed $r$ is a matrix of RLWE ciphertexts in
  RLWE$^{k_0 \times k_1}$, such that $k_0 = \left\lceil s \mathbin{/} a
  \right\rceil$ is the number of RAMs per discriminator, and $k_1 =
  \left\lceil\frac{l2^a}{N} \right\rceil$ is the number of RLWE ciphertexts
  needed to represent a list of $l$ RAMs.
  Let $\pi_r: \text{RGSW}^s \mapsto \text{RGSW}^s$ be a pseudo-random
  permutation map for RGSW ciphertexts deterministic on the value of $r$,
  Algorithms~\ref{alg:he_train} and~\ref{alg:homwiseval} present the training
  and inference on $\mathcal{H}$.

\end{definition}

We start the Integer training process by initializing with zeros the matrix of
RLWE ciphertexts representing the model, and we go over each of the samples in
the training set. We apply the random permutation map $\pi_r$ and partition its
result into $k_0$ vectors of $a$ elements each. These vectors are concatenated
with the bits representing the respective label of the sample, producing a
vector $b$ of $(a + \log_2{l})$ RGSW samples. Then, we use the IVP procedure
over $b$ and an RLWE sample with value 1, which will produce the encrypted
single-valued LUT with value 1 at the position determined by $b$. Finally, this
LUT is added to the model.

\begin{algorithm}[ht]

  \caption{Homomorphic training of the Integer WiSARD model}
  \label{alg:he_train}%
  \SetKwInOut{Input}{Input}
  \SetKwInOut{Output}{Output}
  \Input{
    a Homomorphic WiSARD model $\mathcal{H}_{(s, l, a, r)}$ as defined in
    Definition~\ref{def:homwisard}
  }
  \Input{
    a training set $T$ of size $n$ and its list of labels $L$ encrypted bit by
    bit in RGSW samples
  }
  \Output{trained Homomorphic Integer WiSARD model}
  
  \BlankLine
  $\mathcal{H} \gets \{0\};
  $ $r \stackrel{\mathrm{U}}{\gets} \mathbb{Z};$ \\
  $k_0 \gets \lceil s / a \rceil$; $k_1 \gets \lceil l2^a/N \rceil$ \\
  \For{$i \gets 0$ \KwTo $n - 1$}{
    $t \gets \pi_r(T_i) $ \\
    $u \gets L_i $ \\
    \For{$j \in [\![0, k_0)$}{
      $b \gets [u_0, u_{1}, \dots, u_{\lceil \log_2{l} \rceil - 1},
        t_j, t_{j+1}, \dots, t_{j + a - 1}]$\\ 
      $h \gets \mathsf{IVP}(b, (0,1))$ \\
      \For{$d \in [\![0, k_1)$}{
        $\mathcal{H}[j][d] \gets \mathcal{H}[j][d] + h_d$ \\
      }
    }
  }
  \Return $\mathcal{H}$

\end{algorithm}





\subsection{Model Activation}

Once the Integer training is finished, we start the model activation phase. The
goal at this phase is to apply some activation function $f_{act}$ over the
values of the Integer WiSARD model. There are a few different approaches to
implementing this procedure, which we describe in this section and summarize in
Table~\ref{tab:actcomp}.

\begin{table}[ht]
  \begin{tabular}{r|ccc}
    \hline
    approach & cost
      & \begin{tabular}[c]{c}
        model \\ privacy
      \end{tabular}
      & \begin{tabular}[c]{c}
        continuous \\ learning~\cite{parisi:2019}
      \end{tabular} \\
    \hline
    \textsc{PD-act} & very low 
      & {{\fontencoding{U}\fontfamily{futs}\selectfont\char 49\relax}} 
      & {\checkmark} \\
    \textsc{OTF-act} & high (eval. time) & {\checkmark} & {\checkmark} \\
    \textsc{FM-act}  & amortized low & {\checkmark} & {\xmark} \\
    \hline
  \end{tabular}
  \caption{Comparison between activation approaches.}
  \label{tab:actcomp}
\end{table}


\subsubsection{Post-Decryption Activation (\textsc{PD-act})}

It is very common to have nonlinear operations at the end of an NN inference
process. The most notable example is the use of ARGMAX. When homomorphically
evaluating the inference process, a common strategy to deal with these
non-linear operations is to send all their inputs to the client, who decrypts
them and calculates the operation on the plaintext. This avoids the complexity
and computational cost of evaluating non-linear operations at the cost of
exposing the output scores of the inference to the client. 

The same principle can be adopted for the model activation step of Homomorphic
WiSARDs. One can perform the inference process directly over the Homomorphic
Integer WiSARD model at the server, retrieve the encrypted score of every LUT,
decrypt them individually, and finish the model activation by performing
$f_{act}$, score addition, and ARGMAX on the client. This is the most simple
and inexpensive way of performing model activation, but it comes with two
downsides:

\begin{itemize}

  \item Output size: Models are composed of many RAMs. In our experiments, we
    have up to 6860 RAMs, and each would produce an encrypted score. At first,
    this would require many LWE ciphertexts, which in the case of MNIST would
    use 26.8 MiB of memory. To avoid large result arrays, we can use packing
    key switch procedure~\cite{chillotti:2020} to pack all scores in a single
    RLWE sample, reducing the usage to 131 KiB.

  \item Model privacy: Revealing the individual RAM activation results during
    the evaluation phase enables the client to learn information about the
    model, that could eventually be used to reconstruct it. While treating this
    problem is not within the scope of our work, we note it is possible to use
    simple masking techniques depending on the adopted activation function.

\end{itemize}


\subsubsection{On-the-Fly Activation (\textsc{OTF-act})}

A second approach for performing the activation is to use TFHE's programmable
bootstrapping (PBS) to evaluate $f_{act}$ over the RAM scores of inference. In
this case, one would also perform the inference process directly over the
Homomorphic Integer WiSARD model, and use PBS's to perform $f_{act}$ over the
scores, which could then be added to calculate the score of each discriminator.
At this point, a sequence of PBS's can be used to evaluate the encrypted ARGMAX
function, and only the inferred class is sent to the client. The main advantage
of this approach is fully preserving model privacy as nothing is learned apart
from the prediction. It also doesn't add any cost to the training process. The
main downside, on the other hand, is the cost of performing several PBS's at
every inference, which may be acceptable or not, depending on the context. 


\subsubsection{Full Model Activation (\textsc{FM-act})}

Finally, the third approach for performing the activation consists of using the
PBS to evaluate $f_{act}$ over each element of each RAM of the entire model.
The procedure itself is significantly more expensive than the other approaches,
but it fully protects model privacy and only needs to run once per model, not
affecting inference time. Since the activation is done earlier in the circuit
(right after training, instead of after the inference look-up), it may also
accelerate the training and inference process by allowing them to use smaller
HE parameters (since the noise of the model resets with the bootstrapping). The
main downside of this approach is that, once the model is activated, it loses
some of the capabilities of the Integer WiSARD model. For example, in a
distributed or federated learning scenario, multiple Integer WiSARD models can
be merged by simply adding their RAMs. Once the model is activated, the merge
becomes dependent on the activation function, which often results in non-linear
operations (\emph{e.g.}, when using binarization) or approximations
(\emph{e.g.}, when using threshold activation). This approach is also not
adequate for continuous learning, as it requires the model to be frequently updated.


\subsection{Homomorphic Inference}

The homomorphic evaluation of the inference depends on the approach chosen for
model activation. Algorithm~\ref{alg:homwiseval} shows the inference
considering these choices. The algorithm goes through every RAM in the model,
represented as a vector of RLWE ciphertexts in a line of the model matrix
$\mathcal{H}$. The vector $b$ is the vector of RGSW ciphertexts that serves as
input for Vertical Packing. Its first $\log_2(l)$ bits represent the label, as
in the training, while the remaining are part of the encrypted (and now
permuted) input. We run the VP for all possible values of labels, accumulating
the scores for each class in the vector $\hat{u}$ or saving them in the matrix
$u$ for post-decryption activation. Finally, we finish the algorithm depending
on the choice of model activation method. Supposing the model was already
activated following \textsc{FM-act}, we just need to run the homomorphic
evaluation of the \textsc{ArgMax} function. In the case of \textsc{OTF-act},
the function \textsc{Activate} evaluates the activation function over each
result of VP using programmable bootstrapping. In the \textsc{PD-act} approach,
we just pack everything in a single RLWE ciphertext. Masking techniques are
optional and dependent on the activation function but could also be applied at
this step. Notice that the bits of the label are provided to the VP as
cleartext (we are testing every possible label, there's no secret information
on them), which makes the computation of the VP much faster than the IVP in the
training. On the other hand, we have to run it for every label, which
ultimately leads to similar performance between training and inference. 


\begin{algorithm}[ht]

  \caption{Homomorphic inference}
  \label{alg:homwiseval}%
  \SetKwInOut{Input}{Input}
  \SetKwInOut{Output}{Output}
  \Input{a Homomorphic WiSARD Model $\mathcal{H}_{(s, l, a, r)}$}
  \Input{
    model activation approach
    \textsc{FM-act}, \textsc{OTF-act}, or \textsc{PD-act}
  }
  \Input{a sample $t$ encrypted bit by bit in RGSW ciphertexts}
  \Output{
    classification of $t$ in case of \textsc{FM-act} and \textsc{OTF-act},
    scores of $t$ for every class in case of \textsc{PD-act}
  }
  
  \BlankLine
  $k_0 \gets \lceil s / a \rceil$ \\
  $t'\gets \pi_r(t)$ \\
  $u \gets \{0\}$ \\
  \For{$j \in [\![0,k_0)$}{
    \For{$i \in [\![0,l)$}{
      Let $\hat{l}$ be the bit decomposition of $i$ s.t. $i =
        \sum_{k=0}^ {\lceil \log_2{l} \rceil - 1} \hat{l}_k2^k$ \\
      $b \gets [\hat{l}_0, \hat{l}_{1}, \dots, \hat{l}_{\lceil \log_2{l}
        \rceil - 1}, t_j, t_{j+1}, \dots, t_{j + a - 1}]$ \\ 
      \uIf{\textsc{FM-act}}{
        $\hat{u}_{i} \gets \hat{u}_{i} + \mathsf{VP}(b, \mathcal{H}[j])$
      }\uElseIf{\textsc{OTF-act}}{
        $\hat{u}_{i} \gets \hat{u}_{i} +
          \textsc{Activate}(f_{act},\mathsf{VP}(b, \mathcal{H}[j]))$
      }\uElseIf{\textsc{PD-act}}{
        $u_{i,j} \gets \mathsf{VP}(b, \mathcal{H}[j])$
      }
    }
  }
  \uIf{\textsc{PD-act}}{
    $\hat{u} \gets \textsc{Mask}(u)$ \\
    \Return \textsc{PackingKeySwitching}$(\hat{u})$
  }\uElse{
    \Return \textsc{EncryptedArgMax}$(\hat{u})$
  }
  
\end{algorithm}




\subsection{Additional Techniques}

Our evaluation model enables the easy implementation of several additional
techniques that are commonly needed for privacy-preserving neural network
training and inference. This section discusses some of them.


\subsubsection{Dataset Balancing}\label{sec:hombalance}

In a classification problem with $l$ classes, a dataset set with $n$ samples is
considered balanced if the number of samples for each class is close to $\lceil
n/l \rfloor$. While small unbalances are usually not an issue, highly
unbalanced datasets are a major problem in neural network training, often
leading to skewed models that may even provide artificially high accuracy
without solving the actual classification problem~\cite{alam:2022}. The most
straightforward way of solving balancing problems is to use data augmentation
techniques on samples of underrepresented classes. Another approach is to
adjust the learning rate of the training to correct for the unbalances. For
WNNs, the learning rate would be a factor scaling the impact of every sample
when programming the RAM. More concretely, in line~\ref{line:add} of
Algorithm~\ref{alg:Iwistrain}, our learning rate is the value 1, which is the
same for all samples. We could add values different from one depending on the
class of the input samples. By increasing the value for underrepresented
classes, we would be correcting for the dataset unbalance. 

Both of these approaches, as well as most of the existing ones for typical
neural networks, are problematic for some scenarios of privacy-preserving
machine learning. Specifically, to adjust the learning rate one needs to
pre-process the data set before starting the training, which may not be
possible in scenarios such as distributed or federated learning. It also
requires the data set to be fixed before starting the training, which prevents
techniques such as continuous learning~\cite{parisi:2019}. Balancing through
data augmentation does not present these issues, but it is significantly more
expensive to homomorphically evaluate. Considering our Integer WiSARD model is
a purely linear algorithm, we present a simple and FHE-friendly alternative for
balancing. During training, we use a small IVP over the bits of each sample's
label to create an RLWE ciphertext counting the total number of occurrences of
each class. This data is then sent to the model activation phase, which will
now apply both the activation function and a rescaling according to the
encrypted counting of classes. For the \textsc{PD-act} this is a trivial
procedure, as it can just decrypt the RLWE encrypting the counters and apply
the rescaling on cleartext. For encrypted activations, the process is also
reasonably straightforward, as the programmable bootstrapping can also be used
to apply multivariate functions. More specifically, as introduced
in~\cite{guimaraes:2021} and pointed out in~\cite{chillotti:2021}, the LUT
representing the function to be evaluated by a PBS can be dynamically created
using encrypted data, which allows both function composition and multivariate
evaluation. 


\subsubsection{Federated Learning}

There are many scenarios where the privacy of multiple parties needs to be
simultaneously protected during training or inference. The main example is
federated learning~\cite{zhang:2021}, a case in which multiple independent
(federated) entities collaborate to train a single model together. Each entity
has its own privacy concerns and does not want to share input data with the
others. This requires the use of techniques such as
threshold~\cite{asharov:2012} or multi-key homomorphic
encryption~\cite{chen:2019}, which allows computation to be done over data
encrypted with different or jointly generated keys. Decryption then becomes a
multi-party protocol, in which all involved parties are needed (hence, no data
becomes public without the approval of all involved parties). Even more common
use cases may have similar requirements. In an encrypted inference running in
the public cloud, the owner of the input data and the owner of the model may be
different parties interested in protecting their data (input and model) both
from each other as well as from the cloud provider. 

Most of the existing literature on encrypted NN training and inference does not
consider these aspects, as the techniques required to address them may
introduce a significant computation overhead. Even having both the model and
input sample simultaneously encrypted might already be prohibitive for some
approaches. Homomorphic WiSARDs present many advantages in this regard.
Firstly, having the model encrypted only has a minor impact on the inference
performance. Specifically, our inference could be 2 to 4 times faster if the
model was unencrypted, whereas evaluation models based on other HE schemes may
require much larger encryption parameters. Secondly, only the performance of
the programmable bootstrapping is affected by the use of threshold HE, whereas
other more predominant procedures, such as the VP, are almost unaffected.
Specifically, we would need to use bootstrapping methods such as
LMK$^+$22~\cite{lee:2022}, which supports multiple parties with a
small computational overhead. A third aspect is in the specific case of
federated learning. For general neural networks, the complexity of merging the
models trained by different federated entities may be a challenging procedure
to evaluate homomorphically, as it usually requires rescaling weights and other
non-linear procedures. The Integer WiSARD model allows this merge to be
performed as a simple addition, which enables one to achieve the same performance as in centralized training (as they are functionally equivalent).


\subsubsection{Input Compression}

Compared to other methods for training and evaluating Neural Networks, one
downside of our evaluation method is the ciphertext expansion of the encrypted
input samples. In HE, ciphertext expansion is defined as the factor given by
dividing the size of a ciphertext by the size of the data it encrypts.

We use RGSW ciphertexts and bit-by-bit encryption to enable the fast evaluation
of the VP and IVP procedures. They are $2\ell$ times larger than typical RLWE
ciphertexts, and encrypting just one bit per ciphertext introduces an
additional $O(N)$ factor to the ciphertext expansion. Conversely, computation
using RGSW ciphertexts generates significantly less noise, thus requiring a
much smaller ciphertext modulus $q$. Our approach also does not rely on
batching multiple input samples for efficient training and inference, which is
an advantage for cases where not all input data is available together at once
(e.g., continuous, distributed, or federated learning applications). All in
all, our expansion factor varies from $128\ell N$ to $256\ell N$, whereas
alternative techniques based on RLWE batched approaches depend on the depth of
the network being evaluated, but are usually much smaller.
LTB$^+$23~\cite{legiest:2023} reports an expansion factor of $2 \cdot 389 / 4
\approx 194$ times on a CNN for MNIST with 4-bit linear quantization. 

There are many established ways of avoiding or minimizing ciphertext expansion
for HE schemes. Notably, one could use trans-ciphering to completely negate
ciphertext expansion. For bit-by-bit RGSW encryption, there are also simpler
approaches such as packing multiple bits in a single RGSW ciphertext, which
allows us to reduce the expansion by a factor of $N$ while only requiring a key
switching to unpack the bits. This process is described for the vertical
packing in GNA$^+$22~\cite{guimaraes:2022a}. An alternative solution
for this problem is to use the other HE schemes and computation approaches
adopted by different NN evaluation methods, which we further discuss in
Section~\ref{sec:future}. 


\section{Experimental Results}

We assess the performance of Homomorphic WiSARDs in terms of latency and
accuracy across three popular machine learning datasets. Our methodology and
comparative approach for these datasets are detailed in this section.


\subsection{Datasets}%
\label{sec:datasets}

We employed three datasets: MNIST~\cite{lecun:2010},
HAM10000~\cite{tschandl:2018}, and Wisconsin Breast
Cancer~\cite{wolberg:1995}.


\subsubsection{MNIST}

The MNIST~\cite{lecun:2010} dataset comprises $70000$ grayscale images of
handwritten digits, each sized at $28 \times 28$ pixels. We follow the standard
train and test split and apply linear quantization over the dataset, going from
8 to 4 bits per pixel. We designed four parameter sets to explore parameter
scale and sample volume trade-offs. The MNIST$_T$ (tiny), MNIST$_S$ (small),
MNIST$_M$ (medium), and MNIST$_L$ (large) are trained over $1000$, $7500$,
$30000$ and $60000$ samples, respectively. 


\subsubsection{HAM10000}

The HAM10000~\cite{tschandl:2018} dataset, also known as Skin Cancer MNIST or
DermaMNIST, consists of $10015$ RGB images representing seven types of skin
conditions. Here, we use a $80\%$-$20\%$ train-test set split and apply linear
quantization, going from 8 to 4 bits per pixel. This dataset is known for
being heavily unbalanced~\cite{alam:2022}, and we use our homomorphic balancing
technique (Section~\ref{sec:hombalance}) to avoid overfitting. The
HAM10000$_S$, HAM10000$_M$, and HAM10000$_L$ sets have varying computational
requirements, and are trained over $1002$, $4006$, and $8012$ samples,
respectively.


\subsubsection{Wisconsin Breast Cancer}

The Wisconsin~Breast~Cancer~\cite{wolberg:1995} dataset comprises $569$ samples,
featuring 30 attributes of breast cell nuclei, categorized into benign and
malignant classes. We employ a $80\%$ train set and $20\%$ test set split,
using min-max scaling and linear quantization to convert the features,
originally encoded as floating-point numbers, into 8-bit integers. For this
dataset, we designed one parameter set, as larger sets yield no improvements.


\subsection{Parameter Sets}

We performed an extensive search over the TFHE parameter space, namely $N$,
$\sigma$, $\ell$, and $\beta$, to identify sets that could potentially improve
performance. 
The parameter sets used
for TFHE, presented in Table~\ref{tab:tfhe-params}, were selected to
accommodate the required plaintext space $\mathbb{Z}_p$ for each model.
The parameter sets of the homomorphic WiSARD models, designed for each dataset,
are described in Table~\ref{tab:wisard-params}.

\begin{table}[ht]
  \begin{tabular}{r|cccccc}
    \hline
    set & $\sigma/q$ & $N$ & $\ell$ & $\beta$ 
      & $\ell_{KS}$ & $\beta_{KS}$ \\
    \hline
    HE$_0$ & \multirow{2}{*}{$1.1 \times 2^{-51}$} 
      & \multirow{2}{*}{$2^{11}$} & $1$ & $2^{23}$ & $2$ & $2^{15}$ \\
    HE$_1$ & & & $2$ & $2^{15}$ & $3$ & $2^{11}$ \\
    \hline
  \end{tabular}
  \caption{
    Parameter sets for TFHE. $\ell_{KS}$ and $\beta_{KS}$ are
    decomposition parameters required by \textsc{PackingKeySwitching}.
  }
  \label{tab:tfhe-params}
\end{table}

\begin{table}[ht]
  \begin{tabular}{r|ccccccc}
    \hline
    \multirow{2}{*}{param set} & \multirow{2}{*}{addr} 
      & \multicolumn{2}{c}{therm} & \multirow{2}{*}{act} 
      & \multirow{2}{*}{thr} & \multicolumn{2}{c}{encrypt} \\
    \cline{3-4} \cline{7-8} & & size & type & & & set & p \\
    \hline
    MNIST$_T$ & $9$ & $4$ & log & b-log & $0$ & HE$_0$ & $2^{8}$ \\
    MNIST$_S$ & $12$ & $4$ & log & b-log & $0$ & HE$_0$ & $2^{10}$ \\
    MNIST$_M$ & $14$ & $4$ & log & b-log & $0$ & HE$_1$ & $2^{12}$ \\
    MNIST$_L$ & $16$ & $4$ & log & b-log & $0$ & HE$_1$ & $2^{13}$ \\
    HAM10000$_S$ & $12$ & $5$ & lin & bin & $0$ & HE$_0$ & $2^{10}$ \\
    HAM10000$_M$ & $14$ & $5$ & lin & bin & $1$ & HE$_1$ & $2^{13}$ \\
    HAM10000$_L$ & $16$ & $5$ & lin & bin & $1$ & HE$_1$ & $2^{13}$ \\
    Wisconsin & $10$ & $5$ & lin & log & $0$ & HE$_0$ & $2^{9}$ \\
    \hline
  \end{tabular}
  \caption{
    Parameter sets for all models, with address sizes, thermometer sizes
    and types, activation functions, thresholds, TFHE parameter sets, and
    plaintext moduli.
  }
  \label{tab:wisard-params}
\end{table}


\subsection{Environment Setup}

We experiment in two execution environments, one being a consumer Desktop
computer with an Intel Core i7-12700k clocked at 5.0 GHz with AVX-512 enabled, 
16 threads, and 64GB of memory, and the other an \texttt{i4i.metal} instance on
AWS with an Intel Xeon Platinum 8375C at 3.5 GHz with 128 threads and 1TB of
memory.


\subsection{Encrypted Training in the Cloud}%
\label{sec:rescloud}

We compare the results of our models with the current state-of-the-art for each
dataset. Accuracy is given as the average of 100 executions. Lines in
\textbf{bold} are the results of this work. Evaluation time is always measured
for the entire test set, following the division defined in
Section~\ref{sec:datasets}. For these experiments, we consider the
\textsc{PD-act} activation approach. 


\subsubsection{MNIST}

Our main comparison baseline is Glyph~\cite{lou:2020}, as it represents the
current state-of-the-art for encrypted training. We present both their fastest
result, obtained with one training iteration (epoch), as well as their
highlighted result for better inference. In both cases, we consider only their
accuracy results without transfer learning, as we consider it a
context-specific optimization that is not in the scope of this work (and which
could also be implemented for WNNs~\cite{milhomem:2018}). It is unclear whether
their reported performance results consider transfer learning, which may be
being used to accelerate the training. Glyph~\cite{lou:2020} runs their
experiments in an Intel Xeon E7-8890 v4 at 3.4GHz with 48 threads and 256GB of
memory. Therefore, when calculating our speedup over their techniques, we
consider both the nominal (raw) values as well as a value adjusted (adj) by the
difference in the number of threads. Notice that other factors, such as
architecture, CPU frequency, and compiler version, also affect this comparison,
but it is difficult to consider these differences without access to their
implementation. Table~\ref{tab:comp-mnist-train} presents the training time and
accuracy comparison for our models against Glyph. While we do not implement models with the same accuracy as Glyph, our implementations are significantly faster while only presenting an accuracy drop of 0.4\% to 5.4\%. 

\begin{table}[ht]
  \begin{tabular}{r|cccc}
    \hline
    \multirow{2}{*}{model} & \multirow{2}{*}{acc} 
      & \multirow{2}{*}{time} & \multicolumn{2}{c}{speedup} \\
    \cline{4-5} & & & raw & adj \\
    \hline
    \textbf{MNIST$_S$} & 91.71\% & 3m28s & 3291.4 & 1234.3 \\
    \textbf{MNIST$_M$} & 93.06\% & 38m18s & 300.6 & 112.7 \\
    \textbf{MNIST$_L$} & 93.76\% & 3h30m & 54.9 & 20.6 \\
    Glyph~\cite{lou:2020} & 94.10\% & 1.5d & 5.4 & 5.4 \\
    Glyph~\cite{lou:2020} & 97.10\% & 8d & 1.0 & 1.0 \\
    \hline
  \end{tabular}
  \caption{Training time comparison for MNIST.}
  \label{tab:comp-mnist-train}
\end{table}



\subsubsection{HAM10000}

As in the MNIST comparison, we also adjusted the speedup for differences in
execution environments between our benchmark and Glyph's. Another important
consideration concerns the balancing of the HAM10000 dataset. As discussed by
ASK$^+$22~\cite{alam:2022}, HAM10000 is heavily unbalanced, which often leads
to skewed models that may present artificially high accuracies. For example,
without balancing, if we split the dataset in its original order (taking the
last 20\% of the samples as the test set), we achieve 83.37\% accuracy because
the model completely overfits to a super-represented class, failing to classify
all others. As we balance the model, our (artificially high) total accuracy
falls significantly, but we are able to better classify some of the underrepresented classes.
Glyph~\cite{lou:2020} does not discuss dataset balancing in their work, so it is unclear whether they addressed this problem.
Table~\ref{tab:comp-ham10000-train} presents the results.
Different from MNIST, we achieved gains both in performance and accuracy for HAM10000.

\begin{table}[ht]
  \begin{tabular}{r|cccc}
    \hline
    \multirow{2}{*}{model} & \multirow{2}{*}{acc} 
      & \multirow{2}{*}{time} & \multicolumn{2}{c}{speedup} \\
    \cline{4-5} & & & raw & adj \\
    \hline
    \textbf{HAM10000$_S$} & 67.85\% & 1m35s & 6720.0 & 2520.0 \\
    \textbf{HAM10000$_M$} & 68.60\% & 13m35s & 746.7 & 280.0 \\
    \textbf{HAM10000$_L$} & 69.85\% & 1h03m & 160.0 & 60.0 \\
    Glyph~\cite{lou:2020} & 69.20\% & 7d & 1.0 & 1.0 \\
    \hline
  \end{tabular}
  \caption{Training time comparison for HAM10000.}
  \label{tab:comp-ham10000-train}
\end{table}



\subsubsection{Wisconsin}

As this is a much smaller dataset than the others, we are able to compare our technique against alternatives using
simpler classification algorithms such as support vector machines (SVM) and
multi-layer perception (MLP). Table~\ref{tab:comp-wisconsin-train} presents the results. MFK$^+$24~\cite{montero:2024} run their
experiments in an Intel i7-11800H CPU at 4.6Ghz with 16 threads whereas
PBL$^+$20~\cite{park:2020} uses an Intel Xeon CPU E5-2660 v3 at 3.3 GHz with 20 threads. Adjusting for differences in the machines, our model runs from 166 to 1163 times faster than the alternatives with less than 1\% accuracy deterioration. 

\begin{table}[ht]
  \begin{tabular}{r|cccc}
    \hline
    \multirow{2}{*}{model} & \multirow{2}{*}{acc} 
      & \multirow{2}{*}{time} & \multicolumn{2}{c}{speedup} \\
    \cline{4-5} & & & raw & adj \\
    \hline
    \textbf{Wisconsin} & 97.30\% & 316ms & 9303.8 & 1163.0 \\
    PBL$^+$20 SVM~\cite{park:2020} & 98.00\% & 5m34s & 8.8 & 7.0 \\
    MFK$^+$24 MLP~\cite{montero:2024} & 98.25\% & 49m & 1.0 & 1.0 \\
    \hline
  \end{tabular}
  \caption{Training time comparison for Wisconsin.}
  \label{tab:comp-wisconsin-train}
\end{table}



\subsection{Small Scale Encrypted Training}%
\label{sec:resdesktop}

To show the practicality of our approach, we conducted benchmarks on a consumer-grade desktop computer, using 1 and 8 threads (each of them assigned to operate on a separate physical core). 
The times for medium and large parameter sets were estimated by measuring the execution times across a smaller sample set.
Tables~\ref{tab:desktop-times}~and~\ref{tab:desktop-memory} show the execution time and memory consumption for each model. Our main point with this comparison is to show how practical small-scale encrypted training can be, as small datasets require just a few hundred megabytes of memory to run on a single thread.  

\begin{table}[ht]
  \begin{tabular}{r|cccc}
    \hline
    \multirow{2}{*}{model} & \multicolumn{2}{c}{1 thread}
      & \multicolumn{2}{c}{8 threads} \\
    \cline{2-5} & train & eval & train & eval \\
    \hline
    MNIST$_T$ & 12m16s & 2h11m & 1m33s & 19m40s \\
    MNIST$_S$ & 2h34m & 2h09m & 19m19s & 18m28s \\
    MNIST$_M$ & 28h55m & 4h05m & 3h41m & 33m42s \\
    MNIST$_L$ & 156h15m & 6h04m & 19h57m & 48m45s \\
    HAM10000$_S$ & 54m13s & 1h24m & 6m50s & 11m27s \\
    HAM10000$_M$ & 9h19m & 2h51m & 1h11m & 21m11s \\
    HAM10000$_L$ & 43h30m & 3h45m & 5h27m & 29m31s \\
    Wisconsin & 11s & 3s & 1436ms & 361ms \\
    \hline
  \end{tabular}
  \caption{Consumer desktop benchmark times.}
  \label{tab:desktop-times}
\end{table}

\begin{table}[ht]
  \begin{tabular}{r|cccc}
    \hline
    \multirow{2}{*}{model} & \multicolumn{2}{c}{1 thread}
      & \multicolumn{2}{c}{8 threads} \\
    \cline{2-5} & train & eval & train & eval \\
    \hline
    MNIST$_T$ & 177MiB & 1.0GiB & 501MiB & 1.1GiB \\
    MNIST$_S$ & 427MiB & 1.1GiB & 2.2GiB & 1.1GiB \\
    MNIST$_M$ & 1.2GiB & 1.8GiB & 7.4GiB & 1.8GiB \\
    MNIST$_L$ & 3.4GiB & 3.9GiB & 25.1GiB & 4.0GiB \\
    HAM10000$_S$ & 638MiB & 1.0GiB & 3.9GiB & 1.1GiB \\
    HAM10000$_M$ & 1.8GiB & 2.3GiB & 13.1GiB & 2.3GiB \\
    HAM10000$_L$ & 6.0GiB & 6.2GiB & 45.4GiB & 6.4GiB \\
    Wisconsin & 132MiB & 188MiB & 146MiB & 192MiB \\
    \hline
  \end{tabular}
  \caption{Consumer desktop memory usage.}
  \label{tab:desktop-memory}
\end{table}


\subsection{Model Activation}

Sections~\ref{sec:rescloud} and~\ref{sec:resdesktop} present results
considering the \textsc{PD-act} activation approach, which is the fastest but
relies on masking techniques to fully preserve model privacy.
Table~\ref{tab:resact} presents estimates for the \textsc{FM-act} and
\textsc{OTF-act} approaches considering the use of the programmable
bootstrapping as implemented in LW23~\cite{liu:2023b}. We note that despite
working over large plaintext spaces, the activation functions \textit{b-log} and $bin$ could be implemented with 9-bit precision for all parameters presented in Table~\ref{tab:wisard-params} with probability of failure $2^{-9}$. The
$log$ activation, needed for the Wisconsin problem, already works with $p =
2^9$.

\begin{table}[ht]
  \begin{tabular}{r|cccc}
    \hline
    \multirow{2}{*}{Model} 
      & \multicolumn{2}{l}{\textsc{FM-act} (hours)}
      & \multicolumn{2}{l}{\textsc{OTF-act} (min.)} \\
    \cline{2-5} & 9-bit & 12-bit & 9-bit & 12-bit \\
    \hline
    MNIST$_T$ & 0.1 & 0.4 & \multirow{8}{*}{3.7} & \multirow{8}{*}{21.3} \\
    MNIST$_S$ & 0.4 & 2.1 & & \\
    MNIST$_M$ & 1.1 & 6.4 & & \\
    MNIST$_L$ & 3.8 & 22.0 & & \\
    HAM10000$_S$ & 0.9 & 5.0 & & \\
    HAM10000$_M$ & 2.8 & 16.4 & & \\
    HAM10000$_L$ & 9.8 & 57.2 & & \\
    Wisconsin & 0.1 & 0.4 & & \\
    \hline
  \end{tabular}
  \caption{
    Estimated execution time of other model activation approaches.
    \textsc{FM-act} is the time for activating the full model given in hours,
    estimated for 64 threads. \textsc{OTF-act} is the time to activate the
    result of each inference given in minutes, estimated for a single thread.
    Both consider the PBS execution time from LW23~\cite{liu:2023b}.
  }
  \label{tab:resact}
\end{table}


\subsection{Encrypted Inference}


Table~\ref{tab:inf} presents the comparison among TFHE-based approaches for
encrypted MNIST inference. For our results, we present both single and
multi-threaded results (in parentheses). We note that even our single-threaded
execution is significantly faster than all other approaches for the same
security level. 


\begin{table}[ht]
  \begin{tabular}{r|cccc}
    \hline
    model & acc & time (s) & sec ($\lambda$) & threads \\
    \hline
    DiNN~\cite{bourse:2018} & 93.71 & 0.49 & 80 & 1 \\
    DiNN~\cite{bourse:2018} & 96.3 & 1.5 & 80 & 1 \\
    \hline
    SFB$^+$23~\cite{stoian:2023} & 92.2 & 31 & 128 & 16 \\
    SFB$^+$23~\cite{stoian:2023} & 96.5 & 77 & 128 & 16 \\
    SHE~\cite{lou:2019} & 99.54 & 9.3 & 128 & 20 \\
    REDsec~\cite{folkerts:2023} & 98 & 12.3 & 128 & 96 \\
    REDsec~\cite{folkerts:2023} & 99 & 18.4 & 128 & 96 \\
    \textbf{MNIST}$_S$ & 91.4  & 0.774 (0.048) & 128 & 1 (128) \\
    \textbf{MNIST}$_M$ & 92.81 & 1.47 (0.067) & 128 & 1 (128) \\
    \textbf{MNIST}$_L$ & 93.43 & 2.184 (0.092) & 128 & 1 (128) \\
    \hline
  \end{tabular}
  \caption{
    Comparison of encrypted inference execution time for TFHE-based approaches
    for the MNIST dataset. The security level ($\lambda$) is expressed in bits.
  }
  \label{tab:inf}
\end{table}






\section{Conclusion}\label{sec:conclusion}
In this paper, we introduced the homomorphic evaluation of WNN training and inference algorithms, as well as supporting procedures, such as homomorphic dataset balancing. While WNNs may not still reach the same accuracy levels as CNNs, we showed they are capable of achieving good accuracy levels in just a few minutes of encrypted training, a considerable advance compared to the several days of computation required by previous works. We also showed they have many other advantages besides performance. They can be easily employed for distributed, federated, and continuous learning applications, scenarios that often require major modifications for other types of networks. There are also many opportunities to further improve their performance and accuracy.

\subsection{Further Improvements}\label{sec:future}

\paragraph{Transfer Learning}
A major opportunity for future research is the adoption of transfer learning techniques for WNNs. Glyph~\cite{lou:2020} used transfer learning to improve their CNN results, reporting an increase of up to $4\%$ in accuracy for HAM10000 and 2\% for MNIST. Recently, PTC24~\cite{panzade:2024} employed transfer learning to fine-tune encrypted image recognition models, obtaining further performance and accuracy improvements over Glyph. Transfer learning techniques have also been successfully used for WNNs. 
MAG$^+$18~\cite{milhomem:2018} achieved an up to 11.2\% accuracy improvement over previous literature using transfer learning on a WiSARD for an image classification problem. Compared to their own WiSARD model without transfer learning, the improvement was more than 40\%.


\paragraph{Multi-shot Learning}

Other techniques that may improve WNN-based models are the multi-shot training
procedures such as backpropagation, utilized in the ULEEN~\cite{susskind:2023}
framework. While individual techniques from their models were also tested into
our models, they did not yield any improvements without this specific training
technique. Given that their models achieve up to 98.5\% accuracy on the MNIST
dataset, integrating this technique could potentially narrow the gap between
encrypted WNNs and conventional state-of-the-art CNNs.


\paragraph{Horizontal Packing}

Horizontal packing~\cite{chillotti:2017} is a technique that complements TFHE
vertical packing by allowing the evaluation of multiple LUTs at once. It is
particularly useful for LUTs with sizes smaller than the HE parameter $N$. For
large datasets such as the MNIST and HAM10000, small RAMs didn't show good
accuracy levels, but they seem to be adequate for smaller problems, such as
Wisconsin. For those, horizontal packing can be employed to allow the
evaluation of RAMs from different discriminators simultaneously. 


\paragraph{WNNs and other encryption schemes}

TFHE excels among other HE schemes by its capabilities of efficiently
evaluating LUTs. However, other schemes, such as
BFV~\cite{brakerski:2012b,fan:2012},
BGV~\cite{brakerski:2012a}, and CKKS~\cite{cheon:2017} are
also capable of doing so. They generally present much higher latency for
performing functional bootstrapping, but their throughput of operations may
even surpass TFHE's~\cite{liu:2023b}. 
Conversely, evaluating larger LUTs with similar performance levels as the
Vertical Packing still seems to be a challenge for them. Nonetheless, it should
be practical to evaluate WNNs using these schemes at least for problems
requiring relatively smaller RAMs (e.g., Wisconsin).

\begin{acks}
This work is partially supported by Santander Bank, the São~Paulo Research Foundation (FAPESP, grants 2013/08293-7 and 2019/12783-6), the National Council for Scientific and Technological Development (CNPq, grants 315399/2023-6 and 404087/2021-3), and the \mbox{European Union} (GA 101096435 CONFIDENTIAL-6G). Views and opinions expressed are however those of the author(s) only and do not necessarily reflect those of the European Union or the European Commission. Neither the European Union nor the European Commission can be held responsible for them.
\end{acks}



\bibliographystyle{ACM-Reference-Format}
\bibliography{references}



\appendix


\end{document}